\documentclass[[12pt,thmsa]{article}
\usepackage{amssymb}

\makeatletter \@addtoreset{equation}{section} \makeatother

\def\ftoday{{\sl {Le \number\day \space\ifcase\month
\or janvier\or f\'evrier\or mars\or avril\or mai \or juin\or
juillet\or ao\^ut\or septembre\or octobre \or novembre \or
d\'ecembre\fi\space \number\year}}}
\def\ptoday{{\sl {\number\day \space de\space \ifcase\month
\or janeiro\or fevereiro\or mar{\c c}o\or abril\or maio \or
junho\or julho\or agosto\or setembro\or outubro \or novembro \or
dezembro\fi\space de\space \number\year}}}
\def\gtoday{{\sl {Den \number\day. \ifcase\month
\or Januar\or Februar\or M\"arz\or April\or Mai \or Juni\or
Juli\or August\or September\or Oktober \or November \or
Dezember\fi\space \number\year}}}
\def\today{{\sl {\ifcase\month
\or January\or February\or March\or April\or May \or June\or
July\or August\or September\or October \or November \or
December\fi \space\number\day,\space
                                            \number\year}}}

\renewcommand{\a}{\alpha}
\renewcommand{\b}{\beta}
\newcommand{\g}{\gamma}           
\renewcommand{\d}{\delta}         
\newcommand{\e}{\varepsilon}

\newcommand{\m}{\mu}
\newcommand{\n}{\nu}
         
\newcommand{\p}{\psi}             
\newcommand{\s}{\sigma}           

           \newcommand{\F}{{\Phi}}
\newcommand{\vf}{{\varphi}}
\newcommand{\XI}{\XI}

\newcommand{\LL}{{\cal L}}

\newcommand{\es}{\\[3mm]}

\newcommand{\sla}{\raise.15ex\hbox{$/$}\kern -.57em}
\newcommand{\Sla}{\raise.15ex\hbox{$/$}\kern -.70em}

\def\Lp{\displaystyle{\biggl(}}
\def\Rp{\displaystyle{\biggr)}}

\newcommand{\lp}{\left(}\newcommand{\rp}{\right)}

\newcommand{\complex}{{\kern .1em {\raise .47ex
\hbox {$\scriptscriptstyle |$}}
    \kern -.4em {\rm C}}}
\newcommand{\real}{{{\rm I} \kern -.19em {\rm R}}}
\newcommand{\rational}{{\kern .1em {\raise .47ex
\hbox{$\scripscriptstyle |$}}
    \kern -.35em {\rm Q}}}
\renewcommand{\natural}{{\vrule height 1.6ex width
.05em depth 0ex \kern -.35em {\rm N}}}

\newcommand{\half}{\dfrac{1}{2}}

\newcommand{\dfrac}[2]{{\displaystyle{\frac{#1}{#2}}}}
\newcommand{\dsum}[2]{\displaystyle{\sum_{#1}^{#2}}}
\newcommand{\dint}{\displaystyle{\int}}

\newcommand{\twiddle}{\lower.9ex\rlap{$\kern -.1em\scriptstyle\sim$}}

\newcommand{\vev}[1]{\left\langle {#1}\right\rangle}

\newcommand{\equ}[1]{(\ref{#1})}
\newcommand{\eq}{\begin{equation}}
\newcommand{\eqn}[1]{\label{#1}\end{equation}}
\newcommand{\eea}{\end{eqnarray}}
\newcommand{\eqa}{\begin{eqnarray}}
\newcommand{\eqan}[1]{\label{#1}\end{eqnarray}}
\newcommand{\ba}{\begin{array}}
\newcommand{\ea}{\end{array}}
\newcommand{\eqac}{\begin{equation}\begin{array}{rcl}}
\newcommand{\eqacn}[1]{\end{array}\label{#1}\end{equation}}

\begin{document}

\hfill{{\normalsize
\begin{tabular}{l}
{\sf hep-th/0312193} \\
{\sf CBPF-NF048/03 }  \\
{\sf UFES-DF-OP2003/5}  \\
 \end{tabular}   }
\vspace{3mm}


\begin{center}

{\LARGE\bf Gauge Fixing of Chern-Simons \es $N$-Extended Supergravity}
\end{center}
\vspace{3mm}

\begin{center}{\Large
Wander G. Ney$^{a,b}$, Olivier Piguet$^{c,}$\footnote{Supported in
part by the Conselho Nacional de Desenvolvimento Cient\'{\i}fico e
Tecnol\'{o}gico CNPq -- Brazil.} and Wesley Spalenza$^{a,1}$ }
\vspace{1mm}

\noindent $^{a}$Centro Brasileiro de Pesquisas F\'{i}sicas - (CBPF) -
Rio de Janeiro, RJ\\
$^{b}$Centro Federal de Educa\c{c}\~{a}o Tecnol\'{o}gica - (CEFET)
- Campos dos Goytacazes, RJ\\
$^{c}$ Universidade Federal do Esp\'{i}rito Santo - (UFES) -
Vit\'{o}ria, ES

\vspace{1mm}

{\tt E-mails: opiguet@yahoo.com, wander@cbpf.br, wesley@cbpf.br}

\end{center}

\begin{abstract}
We treat the $N-$extended supergravity in $2+1$ space-time
dimensions as a
Yang-Mills gauge field with Chern-Simons action associated to the $N$%
-extended Poincar\'{e} supergroup. We fix the gauge of this theory
 within the \ Batalin-Vilkovisky scheme.
\end{abstract}


\section{Introduction}

Since Einstein achieved his theory, gravity is treated\ as a particular
geometry of space-time. The metric is what determines
this geometry which, in the first order formalism,
is described by the vierbein -- or more
precisely the dreibein in the context of the present paper -- and
a Lorentz connection. The dreibein components form a
basis of the tangent vector space, equiped with a Lorentz structure. The
metric is then derived from the dreibein. The Lorentz conection, considered as
an independent field, turns out to be
functional of the dreibein by virtue of the field equations.
In this case the action is that of Einstein-Palatini.
$N$-extended supergravity is a supersymmetric generalization of
gravity with $N$ supersymmety generators. We shall be interested in 
$N$-extended supergravity in 3-dimensional space-time, a general presentation
of which having including conformal N-extended supergravity, been given in~\cite{gates-nishino}. 

 Witten \cite{1} described  gravity in $2+1$ dimensions
as a gauge theory, writing the gravity fields as a Yang-Mills field with a
Chern-Simons action and showing the equivalence of the Chern-Simons and
Einstein-Palatini actions. The gauge group of this theory is that
of  Poincar\'{e}, which may be extended to the de
Sitter or anti-de Sitter groups corresponding  to a
 positive  or negative,
 cosmological constant, respectively.

The equivalence of 3-dimensional gravity with a Chern-simons
theory can be generalized to the case of $N$-extended
supergravity, taking the Poincar\'{e}, de Sitter or anti-de Sitter
supergroup as gauge group~\cite{2,3}. All one needs is an invariant
quadratic form for these supergroups from which one can construct
the Chern-Simons action, the latter being then interpreted as the
$N$-extended supergravity action. The author of~\cite{3} did it
for the anti-de Sitter supergroup considered as the product of two
ortho-simplectic supergroups: $AdS(p,q)=OSP_{+}(p,2;{\Bbb R})$
$\times $ $OSP_{-}(q,2;{\Bbb R})$ (where $p+q=N$ is the total
number of supersymmetry generators). They considered various
limiting cases, in particular the super-Poincar\'e limit of
vanishing cosmological constant.

Our purpose is to implement a convenient gauge fixing of this $N$-extended
super-Chern-Simons theory, taking into account the existence of a local
vector supersymmetry generally associated to diffeomorphism invariance as usual in
such topological theories~\cite{vectorsusy}.
Because of the latter invariance the complete gauge algebra
closes only on-shell. We shall use therefore the Batalin-Vilkovisy version of the BRST
gauge fixing scheme~\cite{10}.
In order to obtain the gauge fixed action in a
concise way we shall use a formalism of extended fields, i.e. superpositions
of forms of all possible degrees \cite{ext-fields,11}.

We choose to work
in the present paper directly in the super-Poincar\'{e} limit of null
cosmological constant. The construction for super-de Sitter with
nonvanishing cosmological
constant is similar~\cite{wander} and will not be explicited out here.

 The plan of the paper is the following. Section \ref{ext-sugrav} reviews
the construction of 3-dimensional $N$-extended supergravity as a
Chern-Simons theory fololowing~\cite{3} and then analyses all the
gauge invariances, putting them together in a BRST operator and
writing the corresponding Slavnov-Taylor identity. The gauge
fixing is performed in Section \ref{gauge-fixing}. The paper ends
with a conclusion section.

\section{$N-$Extended Supergravity}\label{ext-sugrav}

In order to construct an $N$-extended supergravity theory a la
Chern-Simons in 3 dimensioanl space-time, with zero cosmological
constant, we choose the $N$-extended Poincar\'{e} supergroup as a
gauge group, whose Lie superalgebra is\footnote{The ``graded
bracket'' $(\ ,\ )$ is an anticommutator if both entries are odd,
and a commutator, otherwise.}:
\begin{equation}
\begin{array}{ccc}
\lbrack J^{a},J^{b}]=\epsilon ^{abc}J_{c}, & [P^{a},P^{b}]=0, &
[J^{a},P^{b}]=\epsilon ^{abc}P_{c}, \\
\lbrack J^{a},Q_{\alpha }^{I}]=-\dfrac{1}{2}(\gamma ^{a})_{\alpha }^{\beta
}Q_{\beta }^{I}, & [P^{a},Q_{\alpha }^{I}]=0, & \left[ Q_{\alpha
}^{I},Q_{\beta }^{J}\right] =\delta ^{IJ}\left( \gamma _{a}\right) _{\alpha
\beta }P^{a}\ .
\end{array}
\label{2}
\end{equation}
$P^{a}$, $J^{a}$ and $Q_{\alpha }^{I}$  are the generators of
space-time translations, Lorentz transformations and supersymmetry transformations,
respectively, the $Q^I$'s being Majorana spinors.
The indices take the values $a=0,1,2$ (Poincar\'e index), $\alpha =1,2$
(spin index) and $I=1,...,N$ (rigid SO($N$) index).
The tangent space metric is Minkowski of signature is $(-++)$,
the Levi-Civita tensor for the Poincar\'{e} indices $\epsilon^{abc}$ $=$
$\epsilon _{abc}$ is defined by $\epsilon^{123}=1$ and the spin Levi-Civita
 tensor $\epsilon_{\alpha \beta }$ $=$ $\epsilon^{\alpha \beta }$
by $\epsilon^{12}=1$. The latter is used in order to lower and rise
the spin indices as
$u_{\alpha }=\mu ^{\beta }\epsilon _{\beta \alpha }$,
$\;u^{\alpha }=\epsilon^{\alpha \beta }u_{\beta }$.
We also have
$\epsilon ^{\alpha \gamma}\epsilon _{\gamma \beta }=
-\delta _{\beta }^{\alpha }$.
The Dirac matrices $(\gamma _{\alpha }^{a\;\beta })$ are chosen real:
$\gamma^{0}=-i\sigma_{y}$, $\gamma^{1}=\sigma_{z}$, $\gamma^{2}=\sigma_{x}$,
the $\s$'s being the Pauli matrices. In this representation the Majorana spinors
have real components: $u_\a^*=u_\a$ and ${\bar u}^\a$ $\equiv$ $(u^+\g^0)^\a$
$=$ $\e^{\a\b}u_\b$ $=$ $u^\a$.

Let us collect in one array $X_A$ the basis elements of the
superalgebra:
\[
\{X_{A}\}=\{P_{a},\,J_{b},\,Q_\alpha^I\,;\
a,b=0,1,2\,;\ \a=1,2\,;\ I=1,\cdots,N\} \ .
\]
In order to construct a Chern-Simons action,
we need a nondegenerate invariant quadratic form
\eq
\vev{\F_1,\F_2} = g_{ab}\F_1^A\F_2^B\ ,\quad \mbox{with}\ \F_{1,2}= \F^A_{1,2}X_A\ ,
\eqn{quad-form}
invariant under the adjoint action of the superalgebra
\[
\d_A\F_{1,2} = [\F_{1,2},X_A]\ .
\]
such a quadratic form may be derived from the following quadratic
Casimir operator of the algebra (\ref{2}): \eq
C=C^{AB}X_{A}X_{B}=P^{a}J_{a}-\dfrac{1}{4}{Q}^{I\a}Q^{I}_\a .
\eqn{casimir}

Namely:
\begin{equation}
\left( g_{AB}\right) =\dfrac{1}{2}\left( -1\right) ^{\left[ X_{A}\right]
}\left( C_{AB}^{-1}\right) =\left(
\begin{array}{ccc}
0 & \left( \delta _{ab}\right) & 0 \\
\left( \delta _{ab}\right) & 0 & 0 \\
0 & 0 & \left( 2\epsilon _{\alpha \beta }\right)
\end{array}\right) ,
\label{inv-form}\end{equation}
where $\left[ X_{A}\right] =0,1$ if $X_{A}$ is an even, odd generator,
respectively.

Writing the Lie algebra valued Yang-Mills connection as \eq A =
A^{A}X_{A}=e^{a}P_{a}+\omega ^{a}J_{a} +\psi^{I\alpha}
Q_{\alpha}^{I}\ , \label{connection}\end{equation} where $e^{a}$,
$\omega ^{a}$ and $\psi^{\alpha I}$ are the dreibein, spin
connection and gravitino 1-forms, respectively, we can now write
the Chern-Simons as
\begin{equation}\ba{l}
S_{\rm CS}= \half\int \langle A,dA+A^{2}\rangle\ ,\es
\phantom{S_{\rm CS}}= -\int e^{a}\left( d\omega
_{a}+\dfrac{1}{2}\epsilon _{abc}\omega ^{b}\omega ^{c}\right)
+{\psi}_{\alpha }^{I} d\psi^{I\alpha }+\dfrac{1}{2} \omega
^{a}{\psi}^{I\alpha }\psi^{I\beta }\gamma _{a\alpha \beta }\ .
\ea\eqn{chern-simon} the symbol $\vev{\ ,\ }$ denoting the
invariant quadratic form \equ{quad-form}. This action is obviously
invariant under rigid SO($N$) transformations, under which $\p$
transforms as a vector and the remaining fields as scalars. We may
turn this invariance into a local SO($N$) invariance substituting
the derivative $d\psi^I$ through the covariant derivative $D_{{\rm
O}}\psi^I=d\psi ^I+A_{{\rm O}}^{IJ}\psi ^{J}$ where $A_{{\rm
O}}^{IJ}$ is a non dynamical SO($N$) connection.

The fields $A_{{\rm O}}^{IJ}$ then play the role of sources of the
conserved Noether currents $j^{IJ\m}$ $=$
$\e^{\m\n\rho}{\psi}^{I}_{\a\n} \psi^{J\a}_\rho$ of the SO($N$)
symmetry. We may observe that the connection $A_{\rm O}$ cannot be
made a dynamical field for the reason that a quadratic Casimir
operator  such as \equ{casimir} containing the SO($N$) generators
is not invertible, hence does not lead to the invariant quadratic
form necessary for writing a kinetic action. This is in contrast
with what happens in the super-anti-de Sitter case, where the
relevant quadratic Casimir operator is indeed
invertible~\cite{3,wander}.

The first term in \equ{chern-simon} is
the Einstein-Palatini action and the others are the
kinetic term of the gravitino and its interaction with the spin conection.

The equations of motion derived from the action (\ref{chern-simon})
read\footnote{The symbol $\stackrel{*}{=}$ means equal up to an
equation of motion ( ''on shell'').}
\begin{equation}
\begin{array}{ll}
\dfrac{\delta S}{\delta e^{a}}=d\omega _{a}+\dfrac{1}{2}\epsilon _{abc}\omega
^{b}\omega ^{c}\stackrel{*}{=}0\ ,\qquad &(\mbox{Curvature}), \es
\dfrac{\delta S}{\delta \omega ^{a}}=de_{a}+\epsilon _{abc}e^{b}\omega ^{c}+%
\dfrac{1}{2}\psi ^{I\alpha }\psi ^{I\beta }\gamma _{a\alpha \beta }\stackrel{*%
}{=}0\ ,\qquad &(\mbox{Torsion}), \es
\dfrac{\delta S}{\delta \psi _{\alpha }^{I}}=2\left( d\psi ^{I\alpha }-\dfrac{1%
}{2}\omega ^{a}\gamma _{a\beta }^{\alpha }\psi ^{I\beta }
\right) \stackrel{*}{=}0\ .\qquad
&(\mbox{Rarita\,\,\,Schwinger}).
\end{array}
\label{eq-de-mov-torcao}
\end{equation}
We see that the curvature is zero. The torsion equation gives us
the spin connection as a function of the dreibein and the
gravitino. The Rarita Schwinger equation expresses the vanishing
of the gravitino covariant derivative with respect to the gauge
group $SO(1,2)$.
The local symmetries of the action are expressed
as its invariance under the BRST transformations
\begin{equation}
sA=-dC-[A,C],\,\,\,\,\,\,\,\,\,\,sC=-C^{2},\,\,\,\,\,\,\,\,\,\,s^{2}=0\
,
\label{sA=-dC-CA}\end{equation}
where the Faddeev-Popov ghost $C$ is written in the adjoint representation of the
Poincar\'{e} supergroup as:
\[
C=C^{A}X_{A}=c_{T}^{a}P_{a}+c_{L}^{a}J_{a}+c_{S}^{I\alpha }Q_{\alpha }^{I}\ ,
\]
the fields $c_{T}^{a}$, $c_{L}^{a}$, $c_{S}^{I\alpha }$ are the ghosts associated
to space-time translations, Lorentz and supersymmetry transformations,
respectively. They are 0-forms of ghost number 1 by definition.
The ghosts $c_{T}^{a}$, $c_{L}^{a}$ are odd and $c_{S}^{I\alpha }$
is even.
In components, the BRST transformations read
\begin{equation}
\begin{array}{ll}
se^{a}=-dc_{T}^{a}+\epsilon _{\;bc}^{a}\omega ^{c}c_{T}^{b}+\epsilon
_{\;bc}^{a}e^{c}c_{L}^{b}-\gamma _{\alpha \beta }^{a}\psi ^{I\alpha
}c_{S}^{I\beta }\ , \quad &
s\omega ^{a}=-dc_{L}^{a}+\epsilon _{\;bc}^{a}\omega ^{c}c_{L}^{b}\ , \es
s\psi ^{I\alpha }=-dc_{S}^{I\alpha }+\dfrac{1}{2}\gamma _{a\beta
}^{\;\;\;\alpha }\omega ^{a}c_{S}^{I\beta }+\dfrac{1}{2}\gamma _{a\beta
}^{\;\;\;\alpha }\psi ^{I\beta }c_{L}^{a}\ , &\es
sc_{T}^{a}=\epsilon _{\;bc}^{a}c_{T}^{c}c_{L}^{b}-\dfrac{1}{2}\gamma _{\alpha
\beta }^{a}c_{S}^{I\alpha }c_{S}^{I\beta }\ , \quad
sc_{L}^{a}=\dfrac{1}{2}\epsilon _{\;bc}^{a}c_{L}^{c}c_{L}^{b}\ , \quad
&sc_{S}^{I\alpha }=\dfrac{1}{2}\gamma _{a\beta }^{\;\;\;\alpha
}c_{L}^{a}c_{S}^{I\beta }\ .
\end{array}
\label{sA=...p cada campo sPoincare}
\end{equation}
The BRST invariance can be expressed through a Slavnov-Taylor identity. We
introduce the Batalin-Vilkovisky anti-fields $A^{*}$ and $C^{*}$ associated to
$A $ and $C$:
\begin{eqnarray}
A^{*} =\omega ^{*a}P_{a}+e^{*a}J_{a}+\psi ^{*\alpha I}Q_{\alpha }^{I}\ , \quad
C^{*} =c_{L}^{*a}P_{a}+c_{T}^{*a}J_{a}+c_{S}^{*I\alpha }Q_{\alpha }^{I}\ ,
\end{eqnarray}
and an action coupling them to the BRST transformations of $A$ and $C$:
\begin{equation}\ba{l}
S_{\rm ext}=\dint \lp\, \langle A^{*},sA\rangle + \langle
C^{*},sC\rangle  \rp\es \phantom{S_{\rm ext}} =\dint \lp
e^{*}se+\omega ^{*}s\omega +\psi _{\alpha }^{*I}s\psi ^{*\alpha I}
+  c_L^{*}sc_{L} + c_T^{*}sc_{T} +  c_S^{*I\a}sc_{S\a}^I  \,\rp \
.\ea\eqn{ext-action} The exterior field $A^{*}$ is a 2-form of
ghost number $-1$ and $C^{*}$ a 3-form of ghost number $-2$. The
Grassmann parities of the fields $A$, $C$, $A^{*}$ and $C^{*}$\
are determined through their total degree: ghost number plus form
degree. The field is even (commuting) or odd (anticommuting) if
its total degree is even or odd respectively.
The total action
\begin{equation}
S=S_{\rm CS}+S_{\rm ext}
\label{tot-action}\end{equation}
obeys the Slavnov-Taylor identity
\begin{equation}
{\cal S}\left( S\right) =\int \sum_{\varphi =A,C}\dfrac{\delta S}{\delta
\varphi ^{*}}\dfrac{\delta S}{\delta \varphi }=0.
\label{slavnov}\end{equation}
We can work in a more compact way defining an extended
field~\cite{ext-fields} $\tilde{A}$
written as the sum of forms of all possible degrees (in our case, degrees 0
to 3):
\begin{equation}
\tilde{A}=C+A+A^{*}+C^{*}\ ,
\end{equation}
the total degree of $\tilde{A}$ being equal 1. We also define an
extended exterior derivative $\tilde{d}=b+d$, where $b$ is a BRST
type operator such that: $b^{2}=[b,d]=0$ \thinspace and therefore
$\tilde{d}^{2}=0$. The extended zero curvature condition is
\begin{equation}
\tilde{F}=\tilde{d}\tilde{A}+\tilde{A}^{2}=0.
\end{equation}
This condition gives us the $b$ transformation of the extended field
\begin{equation}
b\tilde{A}=-d\tilde{A}-\tilde{A}^{2}\ ,  \label{bAAA}
\end{equation}
and hence of its components $C$, $A$, $C^*$ and $A^*$. The
operator $b$ can be interpreted as the linearized Slavnov-Taylor
operator associated with an action $S(\vf,\vf^*)$:
\begin{equation}
b={\cal S}_{S}=\sum_{\varphi =A,C}\int \dfrac{\delta S}{\delta \varphi ^{*}}%
\dfrac{\delta }{\delta \varphi }+\dfrac{\delta S}{\delta \varphi }\dfrac{\delta
}{\delta \varphi ^{*}}\ ,
\end{equation}
provided the action $S$ is a solution of the equations
\begin{equation}
\begin{array}{l}
\dfrac{\delta S}{\delta \varphi ^{*}}={\cal S}_{S}\varphi =b\varphi\ ,\quad
\dfrac{\delta S}{\delta \varphi }={\cal S}_{S}\varphi ^{*}=b\varphi ^{*}\quad
\lp\vf=C,\,A\rp\ .
\end{array}
\label{eq-for-S}\end{equation} The general solution of the latter
equations is the action \equ{tot-action}. This argument shows that
one could have proceeded in a reversed way as well, namely
beginning with the construction of a nilpotemnt operator $b$
acting on the fields and antifields, and then deriving the action
as a solution of \equ{eq-for-S}. This is the procedure we shall
follow in the next sub section. Note that the fulfilment of
\equ{eq-for-S} automatically ensures~\cite{wander} the validity of
the Slavnov-Taylor identity \equ{slavnov}.

\subsection{Diffeomorphism and Vector SUSY}

As a topological one, our theory must be invariant under the
diffeormorfisms or general coordinates transformations. As we want
to include them in the BRST operator, we treat the infinitesimal
diffeormorfism parameter as a ghost vector $\xi =\xi ^{\mu
}\partial _{\mu }$, the components $\xi_\m$ being odd. The
diffeormorfism transformation
\begin{equation}
\delta _{\rm diff}\,\,\varphi =\LL _{\xi }\varphi \ ,\quad
\delta _{\rm diff}\,\,\varphi^* =\LL _{\xi }\varphi^* \ ,
\end{equation}
where $\LL _{\xi }$ is the Lie derivative associated with the
vector field $\xi $, may thus be added to the BRST operator. We do
it in the extended field formalism, still including the local
vector supersymmetry transformation which uses to accompany
diffeomorphism invariance~\cite{vectorsusy}. The ghost of the
latter is an even vector field $v=v^{\mu }\partial_{\mu }$ of
ghost number 2 which happens to contribute to the BRST
transformation of the diffeomorphism ghost $\xi$. In order to get
this more general BRST operator, we define, as explained in the
end of last subsection, a nilpotent $b$-operator which reads, for
the extended field and the new ghosts:
\begin{equation}
\begin{array}{l}
b\tilde{A}=-d\tilde{A}-\tilde{A}^{2}+\LL _{\xi }\tilde{A}-i_{v}\tilde{A}\ ,
\quad b\xi =\xi ^{2}+v\ , \quad bv=[\xi ,v]\ .
\end{array}
\label{ext-brst}\end{equation}
For each form degree this gives
\begin{equation}
\begin{array}{ll}
bC=-C^{2}+\LL _{\xi }C-i_{v}A\ , \quad
&bA=-dC-[A,C]+\LL _{\xi }A-i_{v}A^{*}\ , \es
bA^{*}=-dA-A^{2}-[A^{*},C]+\LL _{\xi }A^{*}-i_{v}C^{*}\ ,\quad
&bC^{*}=-dA^{*}-[A^{*},A]-[C^{*},C]+\LL _{\xi }C^{*}\ ,\es
b\xi =\xi ^{2}+v\ ,\quad
&bv=[\xi ,v]
\end{array}
\label{ext-brs}\end{equation}
or, in components:
\begin{equation}
\begin{array}{l}
bc_{T}^{a}=\epsilon _{bc}^{a}c_{T}^{c}c_{L}^{b}-\dfrac{1}{2}\gamma _{\alpha
\beta }^{a}c_{S}^{I\alpha }c_{S}^{I\beta }+\LL _{\xi
}c_{T}^{a}-i_{v}e^{a}, \\
bc_{L}^{a}=\dfrac{1}{2}\epsilon _{bc}^{a}c_{L}^{c}c_{L}^{b}+\LL _{\xi
}c_{L}^{a}-i_{v}\omega ^{a}, \\
bc_{S}^{I\alpha }=\dfrac{1}{2}\gamma _{a\beta }^{\alpha
}c_{L}^{a}c_{S}^{I\beta }+\LL _{\xi }c_{S}^{I\alpha }-i_{v}\psi
^{I\alpha }, \\
be^{a}=-dc_{T}^{a}+\epsilon _{bc}^{a}e^{c}c_{L}^{b}+\epsilon
_{\;bc}^{a}\omega ^{c}c_{T}^{b}-\gamma _{\alpha \beta }^{a}\psi ^{I\alpha
}c_{S}^{I\beta }+\LL _{\xi }e^{a}-i_{v}\omega ^{*a}, \\
b\omega ^{a}=-dc_{L}^{a}+\epsilon _{\;bc}^{a}\omega ^{c}c_{L}^{b}+\LL %
_{\xi }\omega ^{a}-i_{v}e^{*a}, \\
b\psi ^{I\alpha }=-dc_{S}^{I\alpha }+\dfrac{1}{2}\gamma _{a\beta }^{\alpha
}\omega ^{a}c_{S}^{I\beta }+\dfrac{1}{2}\gamma _{a\beta }^{\alpha }\psi
^{I\beta }c_{L}^{a}+\LL _{\xi }\psi ^{I\alpha }-i_{v}\psi ^{*I\alpha },
\end{array}
\end{equation}
and,  for the anti-fields:
\begin{equation}
\begin{array}{l}
be^{*a}=-de^{a}+\epsilon _{bc}^{a}e^{c}\omega ^{b}-\dfrac{1}{2}\gamma
_{\alpha \beta }^{a}\psi ^{I\alpha }\psi ^{I\beta } \\
\,\,\,\,\,\,\,\,\,\,\,\,\,\,\,\,+\epsilon
_{\;bc}^{a}e^{*c}c_{L}^{b}+\epsilon _{bc}^{a}\omega ^{*c}c_{T}^{b}-\gamma
_{\alpha \beta }^{a}\psi ^{*I\alpha }c_{S}^{I\beta }+\LL _{\xi
}e^{*a}-i_{v}c_{T}^{*a}, \\
b\omega ^{*a}=-d\omega ^{a}+\dfrac{1}{2}\epsilon _{bc}^{a}\omega ^{c}\omega
^{b}+\epsilon _{\;bc}^{a}\omega ^{*c}c_{L}^{b}+\LL _{\xi }\omega
^{*a}-i_{v}c_{L}^{*a}, \\
b\psi ^{*I\alpha }=-d\psi ^{I\alpha }+\dfrac{1}{2}\gamma _{a\beta }^{\alpha
}\omega ^{a}\psi ^{I\beta }+\dfrac{1}{2}\gamma _{a\beta }^{\alpha }\omega
^{*a}c_{S}^{I\beta } \\
\,\,\,\,\,\,\,\,\,\,\,\,\,\,\,\,\,\,\,\,\,\,+\dfrac{1}{2}\gamma _{a\beta
}^{\;\;\;\alpha }\psi ^{*I\beta }c_{L}^{a}+\LL _{\xi }\psi ^{*I\alpha
}-i_{v}c_{S}^{*I\alpha }, \\
bc_{T}^{*a}=-de^{*a}+\epsilon _{\;bc}^{a}e^{*c}\omega ^{b}+\epsilon
_{bc}^{a}\omega ^{*c}e^{b}-\gamma _{\alpha \beta }^{a}\psi ^{*I\alpha }\psi
^{I\beta } \\
\,\,\,\,\,\,\,\,\,\,\,\,\,\,\,+\epsilon
_{\;bc}^{a}c_{T}^{*c}c_{L}^{b}+\epsilon _{bc}^{a}c_{L}^{*c}c_{T}^{b}-\gamma
_{\alpha \beta }^{a}c_{S}^{*I\alpha }c_{S}^{I\beta }+\LL _{\xi
}c_{T}^{*a}, \\
bc_{L}^{*a}=-d\omega ^{*a}+\epsilon _{\;bc}^{a}\omega ^{*c}\omega
^{b}+\epsilon _{\;bc}^{a}c_{L}^{*c}c_{L}^{b}+\LL _{\xi }c_{L}^{*a}, \\
bc_{S}^{*I\alpha }=-d\psi ^{*I\alpha }+\dfrac{1}{2}\gamma _{a\beta
}^{\;\;\;\alpha }\omega ^{*a}\psi ^{I\beta }+\dfrac{1}{2}\gamma _{a\beta
}^{\;\;\;\alpha }\psi ^{*I\beta }\omega ^{a}+\dfrac{1}{2}\gamma _{a\beta
}^{\;\;\;\alpha }c_{L}^{*a}c_{S}^{I\beta } \\
\,\,\,\,\,\,\,\,\,\,\,\,\,\,\,\,\,\,+\dfrac{1}{2}\gamma _{a\beta
}^{\;\;\;\alpha }\psi ^{*I\beta }c_{L}^{a}+\LL _{\xi }c_{S}^{*I\alpha }.
\end{array}
\end{equation}
Now the operator $b$ can be interpreted as the following
 linearized Slavnov-Taylor
operator associated with an action $S(\varphi ,\varphi ^{*},v,\xi )$:
\begin{equation}
{\cal S}_{S}=\int \sum_{\varphi =A,C}\dfrac{\delta S}{\delta \varphi ^{*}}%
\dfrac{\delta }{\delta \varphi }+\dfrac{\delta S}{\delta \varphi }\dfrac{\delta
}{\delta \varphi ^{*}}+\sum_{u=v,\xi }bu\dfrac{\delta }{\delta u}\ ,
\label{ddd1}
\end{equation}
with the transformations $bu$ explicitly\footnote{This means that
we consider the ghost $\xi$ and $v$ as external fields.} given in
\equ{ext-brs}. Indeed, the equations (\ref{eq-for-S}) are solved
by the action
\begin{equation}
S(\varphi ,\varphi ^{*},\xi ,v)=-\dfrac{1}{2}\int \langle \tilde{A},d\tilde{A}%
+\dfrac{2}{3}\tilde{A}^{2}-\LL _{\xi }\tilde{A}+i_{v}\tilde{A}\rangle .
\label{acao exten}
\end{equation}
The integral of an extended form is defined as the integral of its 3-form
terms. This action yields
\eq\ba{ll}
S(\varphi ,\varphi ^{*},\xi ,v) =\dint \Lp &-\dfrac{1}{2}\langle A,\,dA+\dfrac{2}{3%
}A^{2}\rangle +\langle A^{*},\,-dC-[A,C]+\LL _{\xi }A\rangle \es
&+\langle C^{*},-C^{2}+\LL _{\xi }C-i_{v}A\rangle
-\dfrac{1}{2}\langle A^{*},\,i_{v}A^{*}\rangle \Rp
\ea\eqn{min-action}
\section{Gauge Fixing}\label{gauge-fixing}

The gauge invariant action obtained in the preceding subsection
has still to be gauge fixed.
We shall use the Batalin-Vilkovisky scheme~\cite{10,11}. The total action
will then have the form:
\begin{equation}
S_{}=S_{\rm CS}+S_{\rm ext}+S_{\rm gf}\ .
\end{equation}
In order to determine the gauge fixing part $S_{\rm gf}$, we
choose the Landau gauge condition,
which  necessitates the introduction of a nondynamical background metric
$g_{\mu \nu }$\footnote{The dynamical metric is
represented by the dreibein $e^{a}$}.
In differential form notation, the gauge condition reads
\[
d*A=\sqrt{g}\nabla _{\mu}A^{\mu }d^{3}x\ ,
\]
where $\nabla_\m$ is the covariant derivative with respect to
the background metric. It will be implemented through
a Lautrup-Nakanishia Lagrange multiplier $B$ and
its associated  Faddeev-Popov antighost $\bar{C}$, both
being 0-forms and Lie algebra valued:
\begin{equation}
\begin{array}{l}
\bar{C}=\bar{C}^{A}X_{A}=\bar{c}_{T}^{a}P_{a}+\bar{c}%
_{L}^{a}J_{a}+\bar{c}_{S}^{\alpha I}Q_{\alpha }^{I}\ , \qquad
B=B^{A}X_{A}=B_{T}^{a}P_{a}+B_{L}^{a}J_{a}+B_{S}^{\alpha I}Q_{\alpha }^{I}\ ,
\end{array}
\end{equation}
their BRST transformations being defined by
\begin{equation}
s\bar{C}=B\ ,\quad sB=0\ .
\end{equation}
The total gauge fixed action
$S_{}(\varphi ,\varphi ^{*},\xi ,v,B,\bar{C})$
obeying the Slavnov-Taylor identity
\[\ba{l}
{\cal S}(S_{})=\dsum{\varphi}{}\dint\lp
\dfrac{\delta S_{}}{\delta\varphi^{*}}\dfrac{\delta S_{}}{\delta \varphi}
+ (\xi ^{2}+v)\dfrac{\delta  S_{}}{\delta\xi }
+  [\xi ,v]\dfrac{\delta S_{}}{\delta v}
+ B\dfrac{\delta S_{}}{\delta\bar{C}} \rp = 0\ ,
\ea\]
is given, according to Batalin and Vilkovisky, by
\begin{equation}
S_{}(\varphi ,\varphi ^{*},B,\bar{C})=S(\varphi ,\hat{\varphi}%
^{*})+\int Bd*A\ ,
\end{equation}
where the antifields $\varphi^{*}$ in \equ{min-action} have been replaced by
\[
\hat{\varphi}^{*} =
\varphi ^{*}+\dfrac{\delta \Psi }{\delta \varphi}\ ,
\]
the "Batalin-Vilkovisky fermion" $\Psi$ being a local functional
of ghost number $-1$, chosen here as
\begin{equation}
\Psi =\Psi (\varphi ,\bar{C})=\int d\bar{C}*A=\int A*d\bar{C}.
\label{ferm de B V}
\end{equation}
We thus have
\begin{equation}
\hat{A}^{*}=A^{*}+*d\bar{C}\ ,\quad \hat{C}^{*}=C^{*}\ ,
\end{equation}
and the total gauge fixed action reads
\eq\ba{ll}
S(\varphi ,\varphi ^{*},\xi ,v) =\dint \Lp &-\dfrac{1}{2}\langle A,\,dA+\dfrac{2}{3%
}A^{2}\rangle +\langle \hat{A}^{*},\,-dC-[A,C]+\LL _{\xi }A\rangle
\es
&+\langle C^{*},-C^{2}+\LL _{\xi }C-i_{v}A\rangle -\dfrac{1}{2}\langle
\hat{A}^{*},\,i_{v}\hat{A}^{*}\rangle \Rp \ .
\ea\eqn{tot-action'}
In terms of the components fields $e^{a}$, etc., we have
\eq\ba{l}
S = -\dint \Lp  e^{a}d\omega _{a}+\frac{1}{2}\epsilon _{abc}e^{a}\omega
^{b}\omega ^{c}+\psi _{\alpha }^{I}d \psi ^{I\alpha }-\frac{1}{2}\psi
_{\alpha }^{I}\omega ^{a}\psi ^{I\beta }\gamma _{a\beta }^{\;\;\;\alpha }
\es\quad
 +\int \hat{e}^{*a}\left( -dc_{Ta}+\epsilon
_{abc}e^{c}c_{L}^{b}+\epsilon _{abc}\omega ^{c}c_{T}^{b}-\gamma _{a\alpha
\beta }\psi ^{I\alpha }c_{S}^{I\beta }+\LL _{\xi }e_{a}\right)
\es\quad
 +\hat{\omega}^{*a}\left( -dc_{La}+\epsilon _{abc}\omega ^{c}c_{L}^{b}+%
\LL _{\xi }\omega _{a}\right)
    +\hat{\psi}_{\alpha }^{*I}\left( -d c_{S}^{I\alpha }+\frac{1}{2}%
c_{S}^{I\beta }\omega ^{a}\gamma _{a\beta }^{\;\;\;\alpha }+\frac{1}{2}%
\gamma _{\beta }^{a\alpha }c_{La}\psi ^{I\beta }+\LL _{\xi }\psi
^{I\alpha }\right)  \es\quad
 +\hat{c}_{T}^{*a}\left( \epsilon _{abc}c_{T}^{c}c_{L}^{b}-\frac{1}{2}%
\gamma _{a\alpha \beta }c_{S}^{I\alpha }c_{S}^{I\beta }+\LL _{\xi
}c_{Ta}-i_{v}e_{a}\right)
    +\hat{c}_{L}^{*a}\left( \frac{1}{2}\epsilon _{abc}c_{L}^{c}c_{L}^{b}+%
\LL _{\xi }c_{La}-i_{v}\omega _{a}\right) \es\quad
   +\hat{c}_{S\alpha }^{*I}\left( \frac{1}{2}\gamma _{a\beta }^{\alpha
}c_{L}^{a}c_{S}^{I\beta }+\LL _{\xi }c_{S}^{I\alpha }-i_{v}\psi
^{I\alpha }\right)   \es\quad
 -\frac{1}{2}\hat{\omega}^{*a}i_{v}\hat{e}_{a}^{*}-\frac{1}{2}\hat{e}%
^{*a}i_{v}\hat{\omega}_{a}^{*}-\frac{1}{2}\hat{\psi}_{\alpha }^{*I}i_{v}\hat{%
\psi}^{*\alpha I}
+B_{T}^{a}d*\omega _{a}+B_{L}^{a}d*e_{a}+B_{S\alpha }^{I}d*\psi ^{\alpha
I}  \Rp \ ,
\ea\eqn{comp-action}
where
\begin{equation}
*dB_{\mu \nu }=\frac{1}{2\sqrt{g}}\epsilon _{\mu \nu
}^{\,\,\,\,\,\,\,\,\rho }\partial _{\rho }B\ ,
\end{equation}
and
\begin{equation}
\begin{array}{l}
\hat{e}^{*a}=e^{*a}-*d\bar{c}_{T}^{a}\ , \quad
\hat{\omega}^{*a}=\omega ^{*a}-*d\bar{c}_{L}^{a}\ , \quad
\hat{\psi}^{*I\alpha }=\psi ^{*I\alpha }-*d\bar{c}_{S}^{I\alpha }\, \es
\hat{c}_{T}^{*a}=c_{T}^{*a}\ , \quad
\hat{c}_{L}^{*a}=c_{L}^{*a}\ , \quad
\hat{c}_{S}^{*I\alpha }=c_{S}^{*I\alpha }\ .
\end{array}
\end{equation}

This is a gauge fixed action for the $2+1$ dimension $N-$extended
supergravity.

\subsection{Conclusion}

In this work we have achieved the gauge fixing in the
Batalin-Vilkovisky scheme, of
 the $2+1$ dimensional $N$-extended
supergravity considered as
a topological field theory of the Chern-Simons type, the gauge group
being the  $N$-extended Poincar\'{e} supergroup.
We have used the formalism of extended fields in order to express the
BRST algebra, the Slavnov-Taylor identity and the action in
a compact and manageable way.

It would be interesting to consider the construction in terms
of a super-$BF$
topological theory, with the Lorentz group as a gauge group.
This approach would have the advantage of making  possible a
generalization to higher dimensions, the super $B$-field
being then however subjected
to appropriate constraints~\cite{6}, as in the non-supersymmetric
case(see e.g.~\cite{freidel-krasnov}).

\end{document}